\documentclass[fleqn,usenatbib]{mnras}
\usepackage[utf8]{inputenc}
\usepackage{amsmath}
\usepackage{graphicx}
\usepackage{epstopdf}
\usepackage{threeparttable}
\usepackage{graphicx, color}
\usepackage{textcomp}
\usepackage{bm}
\usepackage{soul}
\usepackage{amssymb}
\usepackage{hyperref}
\usepackage{blindtext}

\newcommand{\eqb}{\begin{eqnarray}}
\newcommand{\eqe}{\end{eqnarray}}
\newcommand{\ergs}{erg~s$^{-1}$}
\newcommand{\rt}{r_{\rm t}}
\newcommand{\mel}{m_{\rm e}}
\newcommand{\gmin}{\gamma_{\min}}

\newcommand{\texp}{t_{\rm exp}}
\newcommand{\eB}{\epsilon_{\rm B}}
\newcommand{\ee}{\epsilon_{\rm e}}
\newcommand{\sth}{\sigma_{\rm T}}
\newcommand{\gw}{\gamma_{\rm w}}
\newcommand{\fermi}{{\it Fermi}\xspace}
\newcommand{\swift}{{\it Swift}\xspace}

\newcommand{\xmm}{{\it XMM-Newton}\xspace}

\newcommand{\cxo}{\hbox{\it Chandra}\xspace}

\newcommand{\src}{{PSR~J2032+4127/MT91 213}\xspace}
\newcommand{\psr}{{PSR~J2032+4127}\xspace}

\title[X-rays from interacting winds]
{X-ray mapping of the stellar wind in the binary PSR~J2032+4127/MT91~213}
\author[Petropoulou et al.]{M. Petropoulou$^1$, G. Vasilopoulos$^2$, I.~M. Christie$^1$, D. Giannios$^1$, M.~J. Coe$^3$\\
$^1$Department of Physics, Purdue University, 525 Northwestern Avenue, West Lafayette, IN, 47907, USA \\
$^2$Max-Planck-Institut f\"ur extraterrestrische Physik,Giessenbachstra{\ss}e, 85748 Garching, Germany\\
$^3$Physics and Astronomy and STAG Research Centre, University of Southampton, Southampton, SO17 1BJ, UK
}

\begin{document}

\date{Received.../Accepted...}

\pagerange{\pageref{firstpage}--\pageref{lastpage}} \pubyear{2017}

\maketitle

\label{firstpage}

\begin{abstract}
\psr is a young and rapidly rotating pulsar on a highly eccentric orbit around the high-mass Be star MT91~213. X-ray monitoring of the binary system over a $\sim$4000~d period with \swift has revealed an increase of the X-ray luminosity which we attribute to the synchrotron emission of the shocked pulsar wind. We use \swift X-ray observations to infer a clumpy stellar wind with $r^{-2}$ density profile and constrain the Lorentz factor of the pulsar wind to $10^5<\gamma_{\rm w}<10^6$. We investigate the effects of an axisymmetric stellar wind with polar gradient on the X-ray emission. Comparison of the X-ray light curve hundreds of days before and after the periastron can be used to explore the polar structure of the wind. 
\end{abstract}


\section{Introduction}
\src is a $\gamma$-ray binary system, discovered by the \textit{Fermi} Large Area Telescope (\textit{Fermi}-LAT) \citep{abdo2009} and associated with the source TeV~J2032+4130 \citep{camilo_2009}. The compact object of the binary is PSR J2032+4127, a young ($\sim 0.2$~Myr) pulsar with  spin-down luminosity $L_{\rm sd} \simeq 1.7 \times 10^{35}$~erg~s$^{-1}$ and spin period $P = 143$~ms yielding a magnetic field strength  of $\sim 2 \times 10^{12}$~G. The companion is most likely a Be star \citep[MT91~213:][]{massey_1991} that belongs to the Cygnus OB2 association at a distance $d=1.33\pm0.60$~kpc \citep{kiminki_2015}. The pulsar is on a highly eccentric orbit ($e\sim 0.93-0.99$) around its companion with an orbital period of $P_{\rm orb} \sim 44-49$~yr \citep{ho_2017}. The long $P_{\rm orb}$ and short spin period make the system an outlier of the $P-P_{\rm orb}$  diagram for Be X-ray binaries \citep{corbet_1984}. However, we note many similarities between this system and PSR~B1259-63/LS~2883  \citep{chernyakova_2014}.
  
X-ray monitoring of the system with \swift has revealed a gradual increase of the X-ray luminosity combined with episodes of flaring activity \citep{ho_2017, li2017}. The enhancement of the X-ray emission is believed to arise from the interaction between the relativistic wind of the pulsar and the outflowing wind of the companion Be star. Within this scenario, \cite{takata_2017} have recently argued that the \swift X-ray light curve (LC) can be explained by either a variation in the momentum ratio of the two winds along the orbit or with a pulsar wind magnetization that depends on the distance from the pulsar.

In this \textit{Letter}, we provide   analytical expressions connecting the X-ray luminosity of the binary system with properties of the stellar wind and the relativistic wind of the pulsar. We assume that the observed X-ray emission results from synchrotron emitting pairs which are accelerated at the termination shock formed in the relativistic wind of the pulsar. We search for deviations from the typical $r^{-2}$ density profile of the companion Be star and also provide model X-ray LCs for scenarios where the stellar wind properties have polar and radial dependences. 

The upcoming periastron, occurring in mid-November 2017, provides a unique opportunity to observe variations in the emission of the system from X-rays up to TeV $\gamma$-rays. Follow-up X-ray observations after the periastron passage will allow us to explore anisotropies in the wind properties of a massive star.

\section{Model setup}\label{sec:model_setup}
We adopt the scenario where the observed X-ray emission from \src is attributed to the synchrotron radiation from relativistic electrons and positrons (i.e., pairs). These are accelerated at the termination shock formed by the interacting winds.

The shock terminates at a distance $\rt$ from the neutron star, which can be derived by balancing the ram pressures of the pulsar and stellar winds \citep[e.g.][]{lipunov_1994, tavani_1994}: 
\eqb 
\rt=\sqrt{\frac{L_{\rm sd}}{4\pi c \rho_{\rm w} v_{\rm r}^2}}\approx \sqrt{\frac{L_{\rm sd}}{4\pi c p_{\rm w}}},
\label{eq:rt}
\eqe 
where $p_{\rm w}=\rho_{\rm w} v_{\rm w}^2$ is the ram pressure of the stellar wind and $v_{\rm r}\approx v_{\rm w}$ is relative velocity of the two winds along the orbit\footnote{The pulsar moves at most with $\sim 100$~km s$^{-1}$ at periastron, whereas $v_{\rm w} \sim 10^3$ km s$^{-1}$.}. The shocked pulsar wind moves with a mildly relativistic speed ($\sim c/2$) and expands on the characteristic timescale:
\eqb 
t_{\rm exp}\sim \frac{2\rt}{c}.
\label{eq:texp}
\eqe  
In general, $p_{\rm w}$ is expected to vary with radial distance, thus leading to changes of $\rt$ and $\texp$ along the pulsar's orbit. The magnetic field of the shocked pulsar wind can also be expressed in terms of $p_{\rm w}$ as follows:
\eqb 
B \approx \sqrt{8\pi \eB p_{\rm w}},
\label{eq:B}
\eqe
where the dimensionless parameter $\eB$ is related to the wind magnetization $\sigma$  \citep{kennel_1984} as $\eB=4\sigma$.  At the termination shock, pairs are expected to accelerate  and obtain a non-thermal energy distribution typically described by a power law with slope $p>2$. The pair injection rate into the shocked pulsar wind region can be written as \citep[e.g.][]{christie_2017}:
\eqb 
Q(\gamma) \simeq \ee \frac{L_{\rm sd} (p-2)}{\gmin^2 \mel c^2}\left(\frac{\gamma}{\gmin} \right)^{-p}, \, \gamma \ge \gmin,
\label{eq:qinj}
\eqe
where $\ee \le 1$ is the fraction of the pulsar's spin-down power transferred to relativistic pairs and $\gmin$ is the minimum Lorentz factor of the  pairs, $\gmin=\gw (p-2)/(p-1)$. Among the free parameters of the model, $\gw$ is the most uncertain one with values in the range $10^2-10^6$ \citep[e.g.][]{montani_2014, porth_2014}.

Henceforth, we assume equipartition between particles and magnetic fields in the downstream region of the shock, i.e., $\ee=\eB=0.5$ \citep[e.g.][]{porth_2014}. Synchrotron photons of energy $\epsilon_{\rm x}$  will be produced by pairs with Lorentz factor:
\eqb 
\gamma_{\rm x} \simeq 3\times 10^6 \ p_{\rm w,-4}^{-1/4}\left(\frac{\epsilon_{\rm x}}{5 \ {\rm keV}}\right)^{1/2}
\left( \frac{\eB}{0.5}\right)^{-1/4},
\label{eq:gx}
\eqe 
where $p_{\rm w}=10^{-4} p_{\rm  w, -4}$~g cm$^{-1}$ s$^{-2}$. As long as the synchrotron cooling timescale of the X-ray emitting pairs  is longer than $\texp$ (slow cooling regime), the X-ray synchrotron luminosity emitted over a frequency range $[\nu_{\rm x1}, \nu_{\rm x2}]$  will be given by: 
\eqb 
L_{\rm x}= \mathcal{C}f_{\rm x} L_{\rm sd}^{3/2} \gmin^{p-2} \ee \eB^{\frac{p+1}{4}} p_{\rm w}^{\frac{p-1}{4}},
\label{eq:Lx}
\eqe 
where $f_{\rm x} = \left(\nu_{\rm x1}^{-\beta+1}-\nu_{\rm x2}^{-\beta+1} \right)/(\beta-1)$, $\beta=(p-1)/2$, and $\mathcal{C}=\sth (8\pi)^{\frac{p+1}{4}}\left( 2\pi \mel c/e\right)^{\frac{3-p}{2}}(p-2)/\left(6\pi \mel c^2\sqrt{4\pi c} \right)$. For the derivation of eq.~(\ref{eq:Lx}), we used the $\delta$-function approximation for the synchrotron emissivity, the relation $N(\gamma)=Q(\gamma)\texp$, and eqs.~(\ref{eq:rt})-(\ref{eq:qinj}).

\section{X-ray observations}\label{sec:obs}
\swift/XRT observations can provide a detailed X-ray LC of the system while available \xmm, \cxo, and NuSTAR observations can be used to derive accurate spectral properties. We note that all available observations, apart from those of NuSTAR, yield compatible spectral properties \citep{ho_2017, li2017}. 

The \swift/XRT LC (up to MJD~58033.4) was produced following the instructions described in the \swift data analysis guide (\url{http://www.swift.ac.uk/analysis/xrt/}). We used {\tt xrtpipeline} to generate the \swift/XRT products, and extracted events by using {\tt xselect} \citep[HEASoft FTOOLS;][]{1995ASPC...77..367B}. The LC was also compared for consistency with the automated \swift/XRT online products \citep{2007A&A...469..379E}. To transform the \swift/XRT rates to unabsorbed luminosities, we assumed a distance of 1.3~kpc and a power-law spectrum with photon index $\Gamma=2$ and column density $N_{\rm H}=7.7\times 10^{21}$~cm$^{-2}$ \citep[see Table~4 in][]{ho_2017}.  
\begin{figure}
 \centering 
 \includegraphics[width=0.48\textwidth,trim={0 20 0 20}]{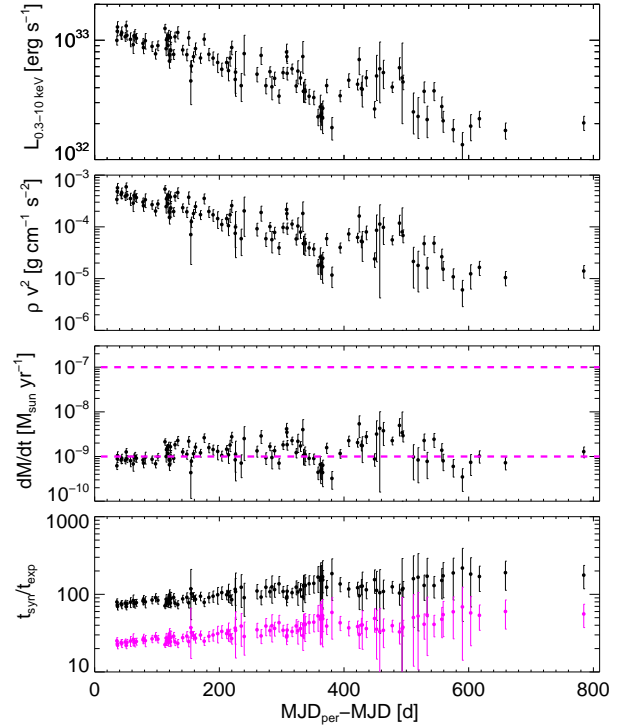}  
\caption{Plotted (from top to bottom) as a function of time till periastron (MJD~58069): the unabsorbed $0.3-10$ keV \swift X-ray LC  of the binary, the ram pressure of the stellar wind, the mass-loss rate of the stellar wind for $v_{\rm w}=10^{3}$~km s$^{-1}$, and the ratio of the synchrotron to expansion timescales for pairs emitting at 0.3 keV (black symbols) and 10 keV (magenta symbols). The horizontal magenta coloured lines indicate the typical range of values for the mass-loss rate of B stars  \protect\citep{martins_2008, krticka_2014}.}  
\label{fig:fig1}
\end{figure}
\section{Results}\label{sec:res}
A transition from the slow cooling to the fast cooling regime can result in a softening of the X-ray spectrum and in the saturation of the system's X-ray luminosity at a fraction of $\ee L_{\rm sd}$, namely:
\eqb 
L_{\rm x, s} & \approx & 2\times 10^{34} \ {\rm erg \ s^{-1}} \ p_{\rm w, -4}^{1/4}  \frac{L_{\rm sd}}{10^{35} {\rm erg \ s^{-1}}} \times \\ \nonumber
 & & \left(\frac{\gw}{10^6}\right)  \left(\frac{\ee}{0.5}\right) \left(\frac{\eB}{0.5}\right)^{1/4} \left(\frac{\epsilon_{\rm x}}{1 \ {\rm keV}}\right)^{-1/2},
\eqe 
where eq.~(\ref{eq:gx}) and $p=3$ were used. In what follows, we assume that pairs emitting in the \swift energy band (0.3-10 keV) are slow cooling (i.e., $t_{\rm cool} > \texp$), since there is no sign of saturation in the \swift data. We benchmark the power-law index of the pair distribution to $p=2\Gamma-1$. The validity of our assumption will also be checked by comparing {\sl a posteriori} the synchrotron cooling and shock expansion timescales.  
\subsection{Spherical stellar wind}\label{sec:res-1}
By matching eq.~(\ref{eq:Lx}) to the unabsorbed \swift X-ray luminosity, we derive $p_{\rm w}$ for any set of the model free parameters (i.e., $\gw$, $\ee$, and $\eB$).  Our results for $\ee=\eB=0.5$ and $\gw=10^6$ are presented in Fig.~\ref{fig:fig1}. 
The top and middle panels show, respectively, the \swift X-ray LC and the inferred ram pressure of the stellar wind, $p_{\rm w} \propto L_{\rm x}^2 \gamma_{\rm w}^{-2} \ee^{-2} \eB^{-2}$. 
The temporal variability of $p_{\rm w}$ may originate from changes in the wind's velocity and/or density. Simulations of line-driven winds from massive stars have shown that both their velocity and density are strongly variable close to the stellar surface \citep[e.g.][]{feldmeier_1995, lobel_2008}. The wind velocity is typically less variable than the density, which may vary more than two orders of magnitude \cite[see Fig.~1][]{bozzo_2016}.  In addition, at the distances of interest (i.e., $r \gg R_{\odot}$), the stellar wind is expected to move with its terminal velocity. As a zeroth order approximation, we thus attribute the changes of $p_{\rm w}$ (middle panel) to changes of the wind's density and assume that $v_{\rm w}$ is constant with radius and equal to its terminal value ($\sim 10^3$~km s$^{-1}$). 

Assuming spherical symmetry, the mass-loss rate of the wind can be estimated as $\dot{M}= 4 \pi r^2 p_{\rm w}/v_{\rm w}$, where $r$ is the separation distance of the binary members. The mass-loss rate fluctuates around an average value of $\sim 10^{-9}$~M$_{\odot}$ yr$^{-1}$ that falls within the value range for B stars  \citep{martins_2008, krticka_2014}. Interestingly, the inferred mass-loss rate is compatible with $\rho_{\rm w} \propto r^{-2}$, while density enhancements giving rise to X-ray flares (top panel in Fig.~\ref{fig:fig1}) can be explained by a clumpy wind \citep[e.g.][]{oskinova_2012}. Because of the scaling  $p_{\rm w}\propto \gw^{-2}$, the inferred mass-loss rate could be as high as $10^{-7}$~M$_{\odot}$ yr$^{-1}$ for a slower pulsar wind with $\gw=10^5$.  Unless the Be star of the binary has an uncommonly high mass-loss rate, then we can set a lower limit on the $\gw$ by requiring that $\langle \dot{M} \rangle \lesssim 10^{-7}$~M$_{\odot}$ yr$^{-1}$:
 \eqb 
 \gamma_{\rm w}\gtrsim 10^5 \left(\frac{0.5}{\ee}\right)\left(\frac{0.5}{\eB}\right)\left(\frac{10^3 \ {\rm km \ s^{-1}}}{v_{\rm w}}\right)^{1/2}.
 \label{eq:gwmin}
 \eqe 
 
The synchrotron cooling timescale of pairs emitting at energy $\epsilon_{\rm x}$ is $t_{\rm syn} \propto \epsilon_{\rm x}^{-1/2} L_{\rm x}^{-3/2} \eB^{3/4} \left(\gamma_{\rm w} \ee\right)^{3/2}$ (see eqs.~(\ref{eq:B}), (\ref{eq:gx})) and is longer than $\texp$ for the adopted parameters (bottom panel in Fig.~\ref{fig:fig1}). Transition to the fast cooling regime during the period of \swift observations would be relevant for $\gw < 10^5$, which would, in turn, imply very high mass-loss rates in contradiction to the expected values for B stars as discussed above (see also eq.~(\ref{eq:gwmin})). Although synchrotron cooling is not relevant for the X-ray emitting pairs during the period of \swift observations, we discuss the possibility of inverse Compton cooling  in section~\ref{sec:disc}. 
 
\begin{figure}
 \centering 
 \includegraphics[width=0.47\textwidth]{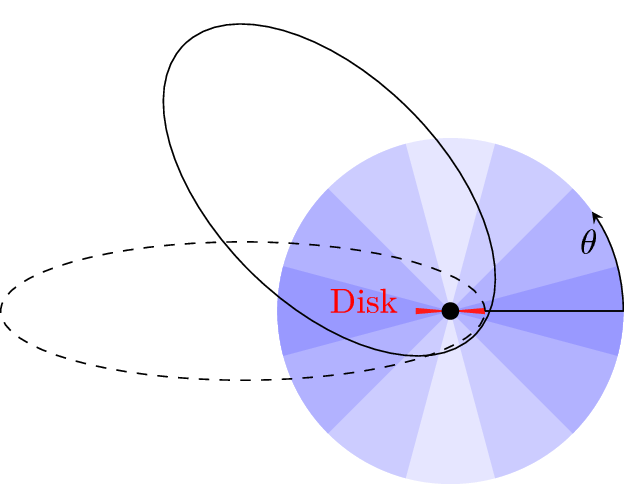}
\caption{Sketch of the Be wind structure. The properties of the wind depend on the distance from the star and on the polar angle $\theta$. The product of the plasma density and velocity becomes lower as we move from the equatorial plane ($\theta=0$) to the pole ($\theta=\pi/2$), as indicated with the colour gradient. The Be disc (red coloured region) is Keplerian and geometrically thin with high plasma density. Two possible orientations of the pulsar's orbit through the wind structure are shown. The pulsar's orbital plane is perpendicular to the Be star's equatorial plane (objects are not in scale).}  
\label{fig:fig2}
\end{figure}

\begin{figure*}
 \centering 
\includegraphics[width=0.98\textwidth, trim=30 0 0 0]{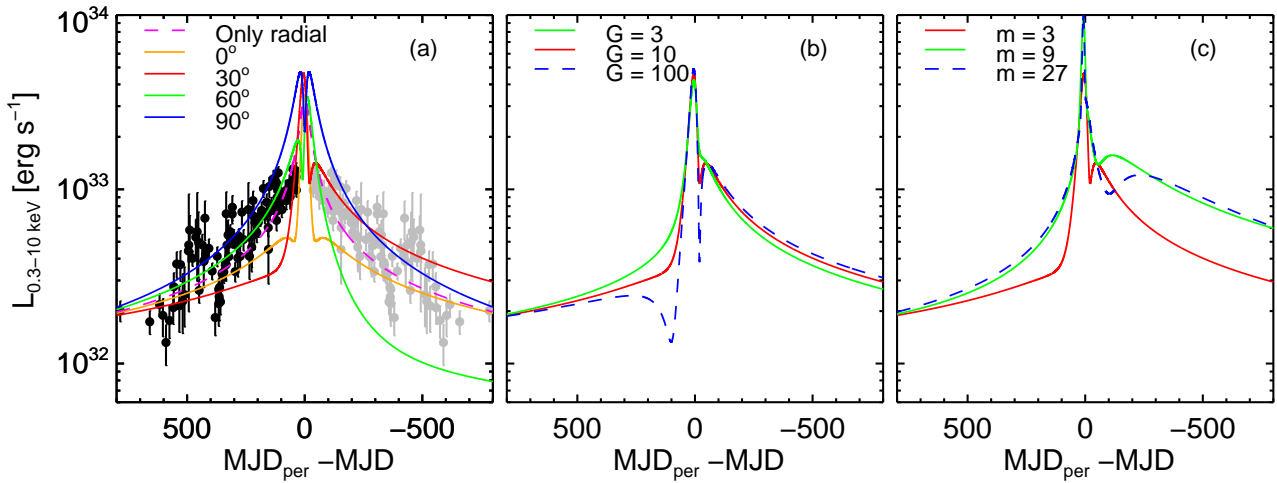}
\caption{X-ray LCs from the interaction of the pulsar wind with different wind structures modelled by eq.~(\ref{eq:prtheta}). (a) We vary 
the angle $\theta_{\rm od}$ (inset legend) for fixed $m=3$ and $G=10$. The LC obtained for a stellar wind without polar dependence (i.e., $G=0$) is also shown for comparison (magenta dashed line). \swift data (MJD 58019.9) and their symmetric ones with respect to the periastron time are over-plotted with black and grey symbols, respectively. (b) We vary the equator-to-pole density contrast $G$ (inset legend)
for fixed $\theta_{\rm od}=30$\textdegree \  and $m=3$. (c) We vary the confinement of wind to the equatorial plane by changing the power-law index $m$ (inset legend) for fixed $\theta_{\rm od}=30$\textdegree \ and $G=10$.  In all panels we used $n=2$ and $p_{\rm w}=10^{-5}$~g cm$^{-1}$ s$^{-2}$ at $\sim 1000$~d before periastron.}  
\label{fig:fig3}
\end{figure*}
\subsection{Axisymmetric stellar wind with polar gradient}\label{sec:res-2}
The extended atmosphere of a Be star can, in general, be divided into two regions: the equatorial (decretion) disc region and the wind region \citep[][]{okazaki_2011}. The former is described by a  geometrically thin Keplerian disc with high plasma density \citep[for a review, see][]{porter_2003}, whereas the wind region is composed of low density plasma moving with $\sim 10^3$ km~s$^{-1}$. 

Here, we consider a toy model for the description of the wind structure. We assume that the stellar wind is axisymmetric (i.e., no azimuthal dependence) and model the ram pressure of the stellar wind as \citep[e.g.][]{petrenz_2000,ignace_2006}:
\eqb 
p_{\rm w}(r, \theta) = p_0 \ r^{-n} \left(1+ G \left|\cos\theta\right|^m\right),
\label{eq:prtheta}
\eqe  
where the distance $r$ is measured from the focus of the ellipse where the Be star is located, the polar angle $\theta \in [0, 2\pi]$ is measured from the Be equator, and $n,m >0$. The constant $G$ is the equator-to-pole density contrast,  the power-law index $m$ determines the confinement of the wind to the equatorial plane, and $p_0$ is a normalization constant. The latter is determined by requiring $p_{\rm w}=10^{-5}$~g cm$^{-1}$ s$^{-2}$ when the binary separation is $4\times 10^{14}$~cm about $1000$~d before periastron. 

We consider the case where the pulsar's orbital plane is perpendicular to the Be's equatorial plane (see Fig.~\ref{fig:fig2}), since we have no knowledge of the system's orientation at the time of writing. The pulsar's orbit can be then expressed in terms of $\theta$ as 
$r_{\rm NS}(\theta)=a(1-e^2)/[1+e\cos\left(\theta - \theta_{\rm od}\right)]$, 
where $e$ is the eccentricity of the orbit and $\theta_{\rm od}$ is the angle between the equator of the star and the orbit's semi-major axis $a$. 

The effect of different wind structures on the X-ray LCs is exemplified in Fig.~\ref{fig:fig3}. For the displayed cases, we keep $n=2$ (see previous section) while varying $\theta_{\rm od}$, $G$, and $m$ separately -- see panels (a) to (c) in Fig.~\ref{fig:fig3}. The LC shape can be symmetric around periastron only for specific orientations; for $\theta_{\rm od}=0$\textdegree, the LC shows a major peak at periastron, while it exhibits two symmetric peaks for $\theta_{\rm od}=90$\textdegree \ as the pulsar crosses the dense equatorial wind twice. All other configurations result in anisotropic LCs. Although the binary separation does not change much close to periastron, the polar variation of the wind becomes stronger for larger $G$ values. This results in LCs with sharp features (dips or peaks) close to the periastron. The duration of these features is determined by the pulsar's speed as it passes through regions with different wind properties along its orbit. Additionally, the $G$ parameter can mimic the effect of different angles between the pulsar's orbital plane and the Be disc plane. The parameter $m$ determines the confinement of the wind to the equatorial plane of the star. Changes in $m$ do not strongly affect the LC close to periastron, but have an impact on the LC shape hundreds of days before or after. A radial wind profile that deviates from the standard one (i.e., $r^{-2}$) may masquerade an axisymmetric stellar wind with polar variations, namely $n=2$ and $m > 0$ (see eq.~(\ref{eq:prtheta})), but cannot explain asymmetric LCs around periastron.

\section{Discussion}\label{sec:disc}
The \swift X-ray data  indicate the presence of a clumpy stellar wind whose density scales, on average, as $\sim r^{-2}$. The latter can explain the gradual increase of the X-ray luminosity as the pulsar approaches periastron, while the X-ray flares are attributed to clumps of dense matter  (section \ref{sec:res-1}). Relativistic hydrodynamic simulations of an inhomogeneous stellar wind interacting with a pulsar wind show that the two-wind interaction region can be perturbed by clumps \citep[see e.g.][]{paredes-fortuny_2014}. These  perturbations may enhance the energy dissipation of the pulsar wind at the termination shock,  strengthen the post-shock magnetic field, or change the direction of motion of the shocked pulsar wind, thus giving rise to X-ray flares \citep[][]{delaCita_2017}. 

Synchrotron cooling is not relevant for X-ray emitting pairs in the $0.3-10$ keV band during the period of \swift observations, as shown in Fig.~\ref{fig:fig1}. This holds also for more energetic pairs that emit in the NuSTAR energy band (i.e., up to 80 keV). It seems therefore unlikely that the softening of the NuSTAR X-ray spectrum in 2016 is caused by synchrotron cooling \citep{li2017}, unless $\gw <10^3$. Inverse Compton (IC) scattering of stellar photons can also be a potential source of pair cooling. For a typical Be star with $T_*=3\times 10^4$~K, pairs radiating at $\epsilon_{\rm x} \propto B \gamma_{\rm x}^2$ will Thomson up-scatter stellar photons from the Rayleigh-Jeans part of the spectrum with $\nu < \nu_{\rm T}\equiv 3 \mel c^2 /h \gamma_{\rm x}$ and energy density $u_{\rm T}\simeq 2\pi k T_* \nu_{\rm T}^3 (10 R_{\odot})^2/(3 c^3 r^2)$. For the same parameters as in Fig.~\ref{fig:fig1}, we find that the shortest IC cooling timescale for pairs emitting at 0.3~keV is $\gtrsim \texp$ at periastron.

A two-wind interaction is expected for the largest part of the pulsar's orbit, whereas the interaction of the pulsar wind with the Be disc may become relevant only close to the periastron \citep[for a detailed study, see][]{takata_2017}. In principle,  the Be disc/pulsar wind interaction can be  studied with eq.~(\ref{eq:prtheta}), since the X-ray luminosity depends on the product of the density and relative velocity of the interacting plasma flows. \cite{takata_2017} derived $\sim 0.2$~g cm$^{-1}$ s$^{-2}$ close to periastron, by adopting a specific model for the Be disc  with  $\rho = 10^{-10}$~g cm$^{-3}$ on the stellar surface and rotational velocity $\sim 10^7$~cm s$^{-1}$ at $r\sim 1$~au (see eqs.~(8)-(9) therein). We obtain similar values using eq.~(\ref{eq:prtheta}) for large $m$ and $G$ values (e.g., $m=27$, $G=100$).

Current H$\alpha$ measurements suggest a Be disc with size $\sim 0.2-0.5$~au \citep{ho_2017}. It is well known that the H$\alpha$ emission only arises in an inner portion of the disc and the actual size of decretion discs can be many times larger. Thus, as the pulsar approaches periastron, it may interact with stellar material. This can be then captured by the neutron star,  if the termination shock lies within its gravitational sphere of influence. This translates to $\rho \gtrsim 3\times 10^{-16}$ g~cm$^{-3}$ at periastron, for $v_{\rm r}=100$~km s$^{-1}$. 
Even in this case, however, a type-I outburst is unlikely to happen \citep[see e.g.][]{okazaki_2001}. Any accretion onto the rapidly rotating neutron star will be halted by the propeller effect \citep{illarionov_1975}, unless the accretion rate is extremely high: $\dot{M}_{\rm a}=(2-20)\times 10^{18}$~g s$^{-1}$ corresponding to $L_{\rm x}=(5-50)\times 10^{38}$~\ergs \ \citep{campana_2002}. 

The pulsar spins down due to  electromagnetic torque $N_{\rm EM}=-\mu^2 \Omega^3/c^3$, where $\Omega$ and $\mu$ are the pulsar's angular spin frequency and magnetic moment, respectively. Close to the periastron,  the pulsar may experience an additional spin-down due to the propeller effect on the plasma ``held'' outside the corotation radius \citep[e.g.][]{illarionov_1975, ghosh_1995}. We adopt the conservative expression of \cite{illarionov_1975} for the propeller torque \citep[for details, see][]{papitto_2012}: $N_{\rm prop}=-\dot{M}_{\rm a}\sqrt{GMR_{\rm m}}\Omega_{\rm K}(R_{\rm m})/\Omega$, where $\Omega_{\rm K}$ is the Keplerian velocity, $R_{\rm m} =\xi \left(\mu^4/2 G M_{\rm NS} \dot{M}_{\rm a}^2\right)^{1/7}$ is the magnetospheric radius, and $\xi\simeq 0.5-1$ \citep[see][and references therein]{chashkina_2017}. The total spin-down torque acting upon the pulsar will be at least doubled, if $\dot{M}_{\rm a}\gtrsim 4\times 10^{17}$~g s$^{-1}$. Assuming that the propeller torque acts upon the pulsar for $\sim 10$~d, we expect a spin-frequency change of $-4\times 10^{-6}$~Hz, i.e. an  anti-glitch \citep{sasmaz_2014}. The predicted effect could be detectable by \fermi--LAT which has already measured a pulsar glitch with $\Delta \nu = 1.9\times 10^{-6}$~Hz \citep{lyne_2015}.

The upcoming periastron passage of \psr offers a unique opportunity to observe the system across the electromagnetic spectrum. In this {\it Letter}, we focused on the non-thermal X-ray emission of the system and showed how it can be used to map the wind properties of the stellar companion. Comparison of the X-ray light curve hundreds of days before and after the periastron will allow us to explore the polar structure of the wind while changes in the pulsar's spin-down rate close to the periastron  may probe the rate of mass captured by the neutron star.

\section*{Acknowledgements} 
We acknowledge support from NASA through the grant NNX17AG21G issued by the Astrophysics Theory Program. We acknowledge the use of publicly available data from the \swift satellite. We thank the \swift team for accepting and carefully scheduling the ToO observations.


\bibliographystyle{mnras} 
\bibliography{psrj032.bib} 

\begin{thebibliography}{}
\makeatletter
\relax
\def\mn@urlcharsother{\let\do\@makeother \do\$\do\&\do\#\do\^\do\_\do\%\do\~}
\def\mn@doi{\begingroup\mn@urlcharsother \@ifnextchar [ {\mn@doi@}
  {\mn@doi@[]}}
\def\mn@doi@[#1]#2{\def\@tempa{#1}\ifx\@tempa\@empty \href
  {http://dx.doi.org/#2} {doi:#2}\else \href {http://dx.doi.org/#2} {#1}\fi
  \endgroup}
\def\mn@eprint#1#2{\mn@eprint@#1:#2::\@nil}
\def\mn@eprint@arXiv#1{\href {http://arxiv.org/abs/#1} {{\tt arXiv:#1}}}
\def\mn@eprint@dblp#1{\href {http://dblp.uni-trier.de/rec/bibtex/#1.xml}
  {dblp:#1}}
\def\mn@eprint@#1:#2:#3:#4\@nil{\def\@tempa {#1}\def\@tempb {#2}\def\@tempc
  {#3}\ifx \@tempc \@empty \let \@tempc \@tempb \let \@tempb \@tempa \fi \ifx
  \@tempb \@empty \def\@tempb {arXiv}\fi \@ifundefined
  {mn@eprint@\@tempb}{\@tempb:\@tempc}{\expandafter \expandafter \csname
  mn@eprint@\@tempb\endcsname \expandafter{\@tempc}}}

\bibitem[\protect\citeauthoryear{{Abdo} et~al.,}{{Abdo}
  et~al.}{2009}]{abdo2009}
{Abdo} A.~A.,  et~al., 2009, \mn@doi [Science] {10.1126/science.1175558}, \href
  {http://adsabs.harvard.edu/abs/2009Sci...325..840A} {325, 840}

\bibitem[\protect\citeauthoryear{{Blackburn}}{{Blackburn}}{1995}]{1995ASPC...77..367B}
{Blackburn} J.~K.,  1995, in {Shaw} R.~A.,  {Payne} H.~E.,   {Hayes} J.~J.~E.,
  eds,  Astronomical Society of the Pacific Conference Series Vol. 77,
  Astronomical Data Analysis Software and Systems IV. p.~367, \url
  {http://adsabs.harvard.edu/abs/1995ASPC...77..367B}

\bibitem[\protect\citeauthoryear{{Bozzo}, {Oskinova}, {Feldmeier}  \&
  {Falanga}}{{Bozzo} et~al.}{2016}]{bozzo_2016}
{Bozzo} E.,  {Oskinova} L.,  {Feldmeier} A.,   {Falanga} M.,  2016, \mn@doi
  [\aap] {10.1051/0004-6361/201628341}, \href
  {http://adsabs.harvard.edu/abs/2016A%26A...589A.102B} {589, A102}

\bibitem[\protect\citeauthoryear{{Camilo} et~al.,}{{Camilo}
  et~al.}{2009}]{camilo_2009}
{Camilo} F.,  et~al., 2009, \mn@doi [\apj] {10.1088/0004-637X/705/1/1}, \href
  {http://adsabs.harvard.edu/abs/2009ApJ...705....1C} {705, 1}

\bibitem[\protect\citeauthoryear{{Campana}, {Stella}, {Israel}, {Moretti},
  {Parmar}  \& {Orlandini}}{{Campana} et~al.}{2002}]{campana_2002}
{Campana} S.,  {Stella} L.,  {Israel} G.~L.,  {Moretti} A.,  {Parmar} A.~N.,
  {Orlandini} M.,  2002, \mn@doi [\apj] {10.1086/343074}, \href
  {http://adsabs.harvard.edu/abs/2002ApJ...580..389C} {580, 389}

\bibitem[\protect\citeauthoryear{{Chashkina}, {Abolmasov}  \&
  {Poutanen}}{{Chashkina} et~al.}{2017}]{chashkina_2017}
{Chashkina} A.,  {Abolmasov} P.,   {Poutanen} J.,  2017, \mn@doi [\mnras]
  {10.1093/mnras/stx1372}, \href
  {http://adsabs.harvard.edu/abs/2017MNRAS.470.2799C} {470, 2799}

\bibitem[\protect\citeauthoryear{{Chernyakova} et~al.,}{{Chernyakova}
  et~al.}{2014}]{chernyakova_2014}
{Chernyakova} M.,  et~al., 2014, \mn@doi [\mnras] {10.1093/mnras/stu021}, \href
  {http://adsabs.harvard.edu/abs/2014MNRAS.439..432C} {439, 432}

\bibitem[\protect\citeauthoryear{{Christie}, {Petropoulou}, {Mimica}  \&
  {Giannios}}{{Christie} et~al.}{2017}]{christie_2017}
{Christie} I.~M.,  {Petropoulou} M.,  {Mimica} P.,   {Giannios} D.,  2017,
  \mn@doi [\mnras] {10.1093/mnrasl/slx017}, \href
  {http://adsabs.harvard.edu/abs/2017MNRAS.468L..26C} {468, L26}

\bibitem[\protect\citeauthoryear{{Corbet}}{{Corbet}}{1984}]{corbet_1984}
{Corbet} R.~H.~D.,  1984, \aap, \href
  {http://adsabs.harvard.edu/abs/1984A%26A...141...91C} {141, 91}

\bibitem[\protect\citeauthoryear{{Evans} et~al.,}{{Evans}
  et~al.}{2007}]{2007A&A...469..379E}
{Evans} P.~A.,  et~al., 2007, \mn@doi [\aap] {10.1051/0004-6361:20077530},
  \href {http://adsabs.harvard.edu/abs/2007A%26A...469..379E} {469, 379}

\bibitem[\protect\citeauthoryear{{Feldmeier}}{{Feldmeier}}{1995}]{feldmeier_1995}
{Feldmeier} A.,  1995, \aap, \href
  {http://adsabs.harvard.edu/abs/1995A%26A...299..523F} {299, 523}

\bibitem[\protect\citeauthoryear{{Ghosh}}{{Ghosh}}{1995}]{ghosh_1995}
{Ghosh} P.,  1995, \mn@doi [\apj] {10.1086/176401}, \href
  {http://adsabs.harvard.edu/abs/1995ApJ...453..411G} {453, 411}

\bibitem[\protect\citeauthoryear{{Ho}, {Ng}, {Lyne}, {Stappers}, {Coe},
  {Halpern}, {Johnson}  \& {Steele}}{{Ho} et~al.}{2017}]{ho_2017}
{Ho} W.~C.~G.,  {Ng} C.-Y.,  {Lyne} A.~G.,  {Stappers} B.~W.,  {Coe} M.~J.,
  {Halpern} J.~P.,  {Johnson} T.~J.,   {Steele} I.~A.,  2017, \mn@doi [\mnras]
  {10.1093/mnras/stw2420}, \href
  {http://adsabs.harvard.edu/abs/2017MNRAS.464.1211H} {464, 1211}

\bibitem[\protect\citeauthoryear{{Ignace} \& {Brimeyer}}{{Ignace} \&
  {Brimeyer}}{2006}]{ignace_2006}
{Ignace} R.,  {Brimeyer} A.,  2006, \mn@doi [\mnras]
  {10.1111/j.1365-2966.2006.10657.x}, \href
  {http://adsabs.harvard.edu/abs/2006MNRAS.371..343I} {371, 343}

\bibitem[\protect\citeauthoryear{{Illarionov} \& {Sunyaev}}{{Illarionov} \&
  {Sunyaev}}{1975}]{illarionov_1975}
{Illarionov} A.~F.,  {Sunyaev} R.~A.,  1975, \aap, \href
  {http://adsabs.harvard.edu/abs/1975A%26A....39..185I} {39, 185}

\bibitem[\protect\citeauthoryear{{Kennel} \& {Coroniti}}{{Kennel} \&
  {Coroniti}}{1984}]{kennel_1984}
{Kennel} C.~F.,  {Coroniti} F.~V.,  1984, \mn@doi [\apj] {10.1086/162357},
  \href {http://adsabs.harvard.edu/abs/1984ApJ...283..710K} {283, 710}

\bibitem[\protect\citeauthoryear{{Kiminki}, {Kobulnicky}, {Vargas {\'A}lvarez},
  {Alexander}  \& {Lundquist}}{{Kiminki} et~al.}{2015}]{kiminki_2015}
{Kiminki} D.~C.,  {Kobulnicky} H.~A.,  {Vargas {\'A}lvarez} C.~A.,  {Alexander}
  M.~J.,   {Lundquist} M.~J.,  2015, \mn@doi [\apj]
  {10.1088/0004-637X/811/2/85}, \href
  {http://adsabs.harvard.edu/abs/2015ApJ...811...85K} {811, 85}

\bibitem[\protect\citeauthoryear{{Krti{\v c}ka}}{{Krti{\v
  c}ka}}{2014}]{krticka_2014}
{Krti{\v c}ka} J.,  2014, \mn@doi [\aap] {10.1051/0004-6361/201321980}, \href
  {http://adsabs.harvard.edu/abs/2014A%26A...564A..70K} {564, A70}

\bibitem[\protect\citeauthoryear{{Li}, {Kong}, {Tam}, {Hou}, {Takata}  \&
  {Hui}}{{Li} et~al.}{2017}]{li2017}
{Li} K.~L.,  {Kong} A.~K.~H.,  {Tam} P.~H.~T.,  {Hou} X.,  {Takata} J.,   {Hui}
  C.~Y.,  2017, \mn@doi [\apj] {10.3847/1538-4357/aa784e}, \href
  {http://adsabs.harvard.edu/abs/2017ApJ...843...85L} {843, 85}

\bibitem[\protect\citeauthoryear{{Lipunov}, {Nazin}, {Osminkin}  \&
  {Prokhorov}}{{Lipunov} et~al.}{1994}]{lipunov_1994}
{Lipunov} V.~M.,  {Nazin} S.~N.,  {Osminkin} E.~Y.,   {Prokhorov} M.~E.,  1994,
  \aap, \href {http://adsabs.harvard.edu/abs/1994A%26A...282...61L} {282, 61}

\bibitem[\protect\citeauthoryear{{Lobel} \& {Blomme}}{{Lobel} \&
  {Blomme}}{2008}]{lobel_2008}
{Lobel} A.,  {Blomme} R.,  2008, \mn@doi [\apj] {10.1086/529129}, \href
  {http://adsabs.harvard.edu/abs/2008ApJ...678..408L} {678, 408}

\bibitem[\protect\citeauthoryear{{Lyne}, {Stappers}, {Keith}, {Ray}, {Kerr},
  {Camilo}  \& {Johnson}}{{Lyne} et~al.}{2015}]{lyne_2015}
{Lyne} A.~G.,  {Stappers} B.~W.,  {Keith} M.~J.,  {Ray} P.~S.,  {Kerr} M.,
  {Camilo} F.,   {Johnson} T.~J.,  2015, \mn@doi [\mnras]
  {10.1093/mnras/stv236}, \href
  {http://adsabs.harvard.edu/abs/2015MNRAS.451..581L} {451, 581}

\bibitem[\protect\citeauthoryear{{Martins}, {Gillessen}, {Eisenhauer},
  {Genzel}, {Ott}  \& {Trippe}}{{Martins} et~al.}{2008}]{martins_2008}
{Martins} F.,  {Gillessen} S.,  {Eisenhauer} F.,  {Genzel} R.,  {Ott} T.,
  {Trippe} S.,  2008, \mn@doi [\apjl] {10.1086/526768}, \href
  {http://adsabs.harvard.edu/abs/2008ApJ...672L.119M} {672, L119}

\bibitem[\protect\citeauthoryear{{Massey} \& {Thompson}}{{Massey} \&
  {Thompson}}{1991}]{massey_1991}
{Massey} P.,  {Thompson} A.~B.,  1991, \mn@doi [\aj] {10.1086/115774}, \href
  {http://adsabs.harvard.edu/abs/1991AJ....101.1408M} {101, 1408}

\bibitem[\protect\citeauthoryear{{Montani} \& {Bernardini}}{{Montani} \&
  {Bernardini}}{2014}]{montani_2014}
{Montani} G.,  {Bernardini} M.~G.,  2014, \mn@doi [Physics Letters B]
  {10.1016/j.physletb.2014.11.029}, \href
  {http://adsabs.harvard.edu/abs/2014PhLB..739..433M} {739, 433}

\bibitem[\protect\citeauthoryear{{Okazaki} \& {Negueruela}}{{Okazaki} \&
  {Negueruela}}{2001}]{okazaki_2001}
{Okazaki} A.~T.,  {Negueruela} I.,  2001, \mn@doi [\aap]
  {10.1051/0004-6361:20011083}, \href
  {http://adsabs.harvard.edu/abs/2001A%26A...377..161O} {377, 161}

\bibitem[\protect\citeauthoryear{{Okazaki}, {Nagataki}, {Naito}, {Kawachi},
  {Hayasaki}, {Owocki}  \& {Takata}}{{Okazaki} et~al.}{2011}]{okazaki_2011}
{Okazaki} A.~T.,  {Nagataki} S.,  {Naito} T.,  {Kawachi} A.,  {Hayasaki} K.,
  {Owocki} S.~P.,   {Takata} J.,  2011, \mn@doi [\pasj]
  {10.1093/pasj/63.4.893}, \href
  {http://adsabs.harvard.edu/abs/2011PASJ...63..893O} {63, 893}

\bibitem[\protect\citeauthoryear{{Oskinova}, {Feldmeier}  \&
  {Kretschmar}}{{Oskinova} et~al.}{2012}]{oskinova_2012}
{Oskinova} L.~M.,  {Feldmeier} A.,   {Kretschmar} P.,  2012, \mn@doi [\mnras]
  {10.1111/j.1365-2966.2012.20507.x}, \href
  {http://adsabs.harvard.edu/abs/2012MNRAS.421.2820O} {421, 2820}

\bibitem[\protect\citeauthoryear{{Papitto}, {Torres}  \& {Rea}}{{Papitto}
  et~al.}{2012}]{papitto_2012}
{Papitto} A.,  {Torres} D.~F.,   {Rea} N.,  2012, \mn@doi [\apj]
  {10.1088/0004-637X/756/2/188}, \href
  {http://adsabs.harvard.edu/abs/2012ApJ...756..188P} {756, 188}

\bibitem[\protect\citeauthoryear{{Paredes-Fortuny}, {Bosch-Ramon}, {Perucho}
  \& {Rib{\'o}}}{{Paredes-Fortuny} et~al.}{2015}]{paredes-fortuny_2014}
{Paredes-Fortuny} X.,  {Bosch-Ramon} V.,  {Perucho} M.,   {Rib{\'o}} M.,  2015,
  \mn@doi [\aap] {10.1051/0004-6361/201424672}, \href
  {http://adsabs.harvard.edu/abs/2015A%26A...574A..77P} {574, A77}

\bibitem[\protect\citeauthoryear{{Petrenz} \& {Puls}}{{Petrenz} \&
  {Puls}}{2000}]{petrenz_2000}
{Petrenz} P.,  {Puls} J.,  2000, \aap, \href
  {http://adsabs.harvard.edu/abs/2000A%26A...358..956P} {358, 956}

\bibitem[\protect\citeauthoryear{{Porter} \& {Rivinius}}{{Porter} \&
  {Rivinius}}{2003}]{porter_2003}
{Porter} J.~M.,  {Rivinius} T.,  2003, \mn@doi [\pasp] {10.1086/378307}, \href
  {http://adsabs.harvard.edu/abs/2003PASP..115.1153P} {115, 1153}

\bibitem[\protect\citeauthoryear{{Porth}, {Komissarov}  \& {Keppens}}{{Porth}
  et~al.}{2014}]{porth_2014}
{Porth} O.,  {Komissarov} S.~S.,   {Keppens} R.,  2014, \mn@doi [\mnras]
  {10.1093/mnras/stt2176}, \href
  {http://adsabs.harvard.edu/abs/2014MNRAS.438..278P} {438, 278}

\bibitem[\protect\citeauthoryear{{Takata}, {Tam}, {Ng}, {Li}, {Kong}, {Hui}  \&
  {Cheng}}{{Takata} et~al.}{2017}]{takata_2017}
{Takata} J.,  {Tam} P.~H.~T.,  {Ng} C.~W.,  {Li} K.~L.,  {Kong} A.~K.~H.,
  {Hui} C.~Y.,   {Cheng} K.~S.,  2017, \mn@doi [\apj]
  {10.3847/1538-4357/aa5c80}, \href
  {http://adsabs.harvard.edu/abs/2017ApJ...836..241T} {836, 241}

\bibitem[\protect\citeauthoryear{{Tavani}, {Arons}  \& {Kaspi}}{{Tavani}
  et~al.}{1994}]{tavani_1994}
{Tavani} M.,  {Arons} J.,   {Kaspi} V.~M.,  1994, \mn@doi [\apjl]
  {10.1086/187542}, \href {http://adsabs.harvard.edu/abs/1994ApJ...433L..37T}
  {433, L37}

\bibitem[\protect\citeauthoryear{{{\c S}a{\c s}maz Mu{\c s}}, {Ayd{\i}n}  \&
  {G{\"o}{\u g}{\"u}{\c s}}}{{{\c S}a{\c s}maz Mu{\c s}}
  et~al.}{2014}]{sasmaz_2014}
{{\c S}a{\c s}maz Mu{\c s}} S.,  {Ayd{\i}n} B.,   {G{\"o}{\u g}{\"u}{\c s}} E.,
   2014, \mn@doi [\mnras] {10.1093/mnras/stu436}, \href
  {http://adsabs.harvard.edu/abs/2014MNRAS.440.2916S} {440, 2916}

\bibitem[\protect\citeauthoryear{{de la Cita}, {Bosch-Ramon},
  {Paredes-Fortuny}, {Khangulyan}  \& {Perucho}}{{de la Cita}
  et~al.}{2017}]{delaCita_2017}
{de la Cita} V.~M.,  {Bosch-Ramon} V.,  {Paredes-Fortuny} X.,  {Khangulyan} D.,
    {Perucho} M.,  2017, \mn@doi [\aap] {10.1051/0004-6361/201629112}, \href
  {http://adsabs.harvard.edu/abs/2017A%26A...598A..13D} {598, A13}

\makeatother
\end{thebibliography}
 
\end{document}